\documentclass[aps,prb,twocolumn,superscriptaddress,groupedaddress,floatfix,showpacs]{revtex4}
\usepackage{amsmath}
\usepackage{amsfonts}
\usepackage{amssymb}
\usepackage{graphicx}
\usepackage{bm}

\begin{document}

\title{Van der Waals forces and electron-electron interactions in two strained graphene layers}
\author{Anand Sharma} 
\email{anand.sharma@uvm.edu}

\author{Peter Harnish} 

\author{Alexander Sylvester} 

\author{Valeri N. Kotov}
\affiliation{Department of Physics, University of Vermont, 82 University Place, Burlington, Vermont 05405, USA}

\author{A. H. Castro Neto}
\altaffiliation{On leave from Department of Physics, Boston University, 590 Commonwealth Avenue, Boston, Massachusetts 02215, USA}
\affiliation{Graphene Research Centre and Department of Physics, National University of Singapore, 2 Science Drive 3, Singapore 117542}

\begin{abstract}

We evaluate the van der Waals (vdW) interaction energy at zero temperature between two undoped strained graphene layers  separated by a finite distance. We consider the following three models for the anisotropic case: (a) where one of the two layers is uniaxially strained, (b) the two layers are strained in the same direction, and (c) one of the layers is strained in the perpendicular direction with respect to the other. We find that for all  three models and   given value of the electron-electron interaction coupling, the vdW interaction energy increases with increasing anisotropy. The effect is most striking for the case when both  layers are strained in the same direction where we observe up to an order of magnitude increase in the strained relative to the unstrained case. We also investigate the effect of  electron electron interaction renormalization in the region of large separation between the strained graphene layers. We find that the many-body renormalization contributions to the correlation energy are non negligible and the vdW interaction energy decreases as a function of increasing distance between the layers due to renormalization of the Fermi velocity, the anisotropy, and the effective interaction. Our analysis can be useful in designing novel graphene-based vdW heterostructures which, in recent times, has seen an upsurge in research activity.

\end{abstract}

\date{\today}


\maketitle

\section{\label{sec:int} Introduction}

\indent Graphene, an atomic-thin sheet of carbon atoms, was isolated from graphite using the micromechanical cleavage technique\cite{novoselov}. It has remarkable mechanical\cite{lee}, electronic\cite{castroneto}, transport\cite{peres} and optical\cite{grigorenko} properties. Since its isolation in 2004, it has drawn a lot of interest due to its unique zero gap electronic band-structure at the Dirac point with chiral and massless linear dispersion of Dirac fermions.\\
\indent In the past several years, the research has rapidly moved on from single to bi-\cite{mccann}, double-\cite{ponomarenko} and multi-layer graphene\cite{nilsson} as they display plethora of intriguing properties due to the inherent chiral symmetry of the underlying bipartite lattice structure. Such multi layer systems are formed by stacking graphene on top of each other and they can be either electronically coupled or decoupled. The spatially separate double-layer graphene\cite{schmidt}, where they are coupled only via the long-range Coulomb interaction, are particularly very fascinating as they exhibit variety of phenomena like the plasmon effects\cite{stauber}, excitonic condensate or frictional Coulomb drag\cite{gorbachev} and van der Waals (vdW) interaction\cite{sernelius1}. It turns out that these physical phenomena are not only useful in probing the interaction effects in such systems but are also important in designing modern technological devices as they can be separated in the order of nanometer scale.\\
\indent Recently the low dimensional vdW heterostructures has attracted a great deal of attention as they provide a stage for new materials to study interesting effects and to realize quantum engineered devices with unprecedented qualities for modern applications\cite{britnell}. The van der Waals force between electrically neutral atoms or molecules arises due to the instantaneous dipole induced by the fluctuating electron cloud around the nucleus and is normally attractive. It is so ubiquitous in nature that its study covers almost all areas of natural sciences and has wide range of utilization among interdisciplinary subjects\cite{langbein}. This long range and weak force plays a very crucial role in the investigation of interaction between the materials\cite{gao}. Thus thorough understanding of vdW interactions and utilizing non-covalent nanomaterials, for instance carbon-based graphene, forms a crucial step towards future technology\cite{bunch}.\\
\indent Graphene as a two-dimensional sheet has extraordinary mechanical strength\cite{lee}, but can also be subjected to strain leading to anisotropic electronic behavior. It is known that the lattice distortion, due to uniaxial strain, in graphene can alter its electronic band structure\cite{pereira1,park,goerbig}, optical\cite{pereira2} as well as magnetic\cite{sharma1} properties and also demonstrates interplay with many-body interactions\cite{sharma2} and electron-plasmon scattering\cite{pellegrino}. It is also possible to open a finite band gap in the electronic spectrum, which would be highly desirable for the semiconductor-based technology\cite{schwierz}, but only at the expense of very large deformations applied along a specific direction\cite{pereira2}. Due to the lack of achieving substantial deformations in graphene, there has not been been much work reported on applying strain in two- or multi-layered graphene.\\
\indent Besides tailoring the electronic band-structure of graphene using strain, the many-body interactions also play an important role in understanding the fundamental physical properties as well as for applications in technical devices. These interactions are normally more pronounced in low dimensional materials and are visible as intrinsic enhancement of Coulomb interactions as well as reduced screening in these materials. In graphene, the low-energy behavior of the Dirac quasiparticles is remarkably modified due to the electron-electron interactions and is evident in terms of the fractional quantum Hall effect\cite{du} and renormalization of the Fermi velocity\cite{elias} among various other striking results\cite{kotov}. In addition to the mono-layer, various interesting phenomena related to many-body interactions were also observed in double-layer graphene\cite{min,chae,sarabadani,dobson1}. \\
\indent Thus it would be captivating to study the effects of the electron-electron interactions along with strain on the vdW interactions between two-layer graphene and this forms the main motivation of this paper. Our aim is to obtain the vdW force acting between two uniaxially strained clean graphene layers at zero temperature and which are separated by a finite distance. In Section~\ref{sec:amd} we introduce the three anisoptropic models and obtain the vdW interaction energy per unit area in each of these cases. We consider an effective Coulomb interaction between the graphene sheets derived with in Random Phase Approximation (RPA). The numerical results, discussed in Section~\ref{sec:numres}, are divided in to two parts. In subsection~\ref{sec:woeei} we examine the variation in the vdW force coefficient as a function of applied uniaxial strain for different values of the strength of effective interaction or coupling constant. And in subsection~\ref{sec:weei} we additionally consider the effect of intra-layer electron-electron interaction on the vdW interaction energy as the distance between the two layers is increased. In Section~\ref{sec:sumcon} we summarize the results and present our conclusions.\\

\section{\label{sec:amd} Anisotropic models and the van der Waals force}

\indent We are interested in evaluating the van der Waals (vdW) interaction energy or the vdW force per unit area at zero temperature (T=0) between two charge neutral strained graphene layers separated by a finite distance. We begin by indicating that for any interacting many electron system, the total energy consists of its kinetic, exchange and correlation energy. In our study, the intra-layer kinetic and the exchange energy are independent of the distance between the two layers and there's no inter-layer exchange energy since we assume that the separation between the two layers are large enough to neglect any overlap between the electronic wavefunctions. Thus the many-body electronic correlation energy, a part of which is the weak long-ranged vdW interaction energy, is the only quantity which depends on the separation between the layers\cite{rydberg,dobson2}. The vdW interaction arises because of the induced dipole-dipole interaction due to the correlated motion of the electrons in both the layers and in the presence of an effective inter-layer Coulomb interaction between the layers.\\

\begin{figure}[ht]
\centering
\includegraphics[width=\columnwidth]{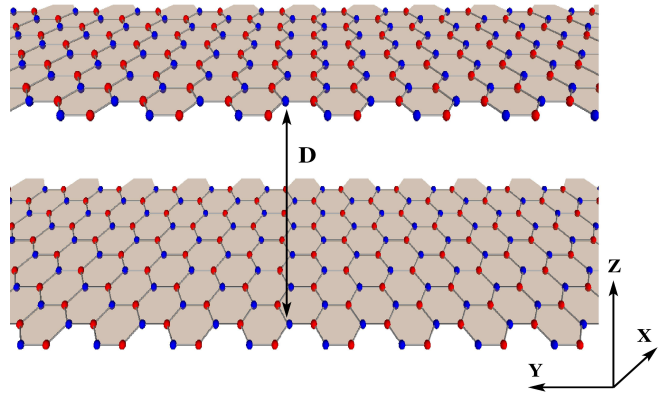}
\caption{\label{fig:1} (Color online) Schematic depicting two undoped and unstrained freely suspended graphene layers separated by a finite distance (D).}
\end{figure}

\indent For a system of two undoped and unstrained freely suspended graphene layers separated by a finite distance (D), as shown in Fig.~\ref{fig:1}, the vdW force per unit area at T=0 was derived in several different ways\cite{sernelius2,dobson3} and a consistent power law dependence on the distance was obtained which varied as D$^{-4}$  in both the non-retarded\cite{sernelius2} and retarded\cite{sernelius3}  regimes. This was due to the fact that in such a system for no doping and zero temperature, the force of attraction resulted from the interaction of the charge-density oscillations within the graphene crystal and there was no characteristic temperature scale. Therefore if one neglected the effects of intra-layer Coulomb interaction then there was no physical quantity which had a dependence on the distance of separation between the graphene layers. Moreover, under that condition the retardation effects were also shown to be practically irrelevant for such a system\cite{gomezsantos}. We would also like to remark that, surprisingly, the above mentioned power law dependence was in sharp contrast to the case of two layers of two-dimensional metals (D$^{-\frac{7}{2}}$) or insulators (D$^{-5}$) and the difference was attributed to the unique bandstructure of graphene\cite{dobson4}. There have also been studies reporting finite temperature\cite{sarabadani,gomezsantos,klimchitskaya} calculations and with doped as well as gapped graphene\cite{drosdoff}.\\

\begin{figure}[htbp]
\centering
\includegraphics[height=6.0cm,width=9.0cm]{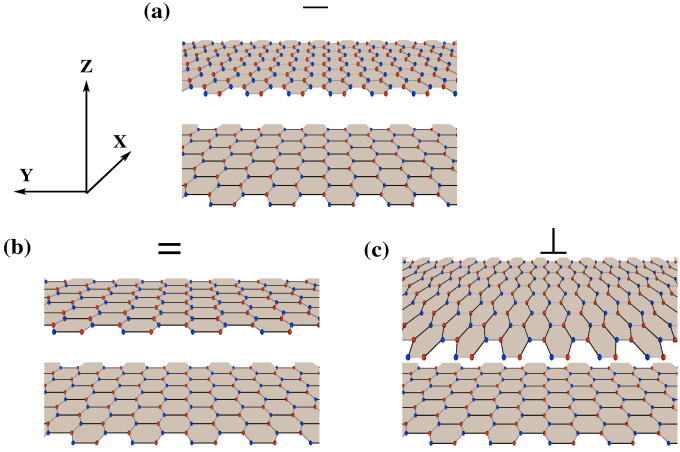}
\caption{\label{fig:2} (Color online) Schematic showing three different anisotropic models; (a) one of the two graphene layers (bottom layer as shown in the figure) is uniaxially ($-$) strained, (b) the two layers are strained in the same ($=$) direction and (c) one of the layers is strained in the perpendicular ($\perp$) direction with respect to the other for a fixed co-ordinate system.}
\end{figure}

\indent In this work we consider three different types of geometries for an applied uniaxial strain on graphene sheets, as shown in Fig.~\ref{fig:2}, with (a) one of the two graphene layers (bottom layer as shown in the figure) is uniaxially ($-$) strained, (b) the two layers are strained in the same ($=$) direction and (c) one of the layers is strained in the perpendicular ($\perp$) direction with respect to the other for a fixed co-ordinate system. Our starting point is the evaluation of the vdW interaction energy at zero temperature. It is known that the result for the interaction energy within the Random Phase Approximation (RPA)~\cite{ren}, which takes into account both intra- and inter-layer screening, is equivalent to the non-retarded version of the well-known Lifshitz approach \cite{lifshitz}. Retardation effects can be easily taken into account  but are known to be negligible in unstrained graphene \cite{gomezsantos}. However we will show later that in the limit of large strain retardation effects have to be taken into account, otherwise the force grows uncontrollably. Thus our starting expression for the interaction energy is:

\begin{eqnarray}{\label{e1}}
\mathcal{E}(D) &=& \frac{\hbar}{(2\pi)^{3}}\iint dq_x dq_y\int_{0}^{\infty} d\omega \nonumber \\
&& \ln\left(1 - \frac{e^{-2qD}\displaystyle V_1({\textbf{q}}) \Pi_{1}(\textbf{q},i\omega)
V_2({\textbf{q}}) \Pi_{2}(\textbf{q},i\omega)}{\displaystyle
\epsilon_{1}(\textbf{q},i\omega)\epsilon_{2}(\textbf{q},i\omega)}\right) \nonumber \\
\end{eqnarray}
where  the dielectric functions in the layers,  labeled $\nu = 1$(top) and $2$(bottom), are given as
 $\epsilon_{\nu}(\textbf{q},i\omega) = 1 - V_{\nu}({\textbf{q}})\Pi_{\nu}(\textbf{q},i\omega)$. Here $V_1({\textbf{q}}) = V_2({\textbf{q}}) = V({\textbf{q}}) = \frac{2\pi e^2}{\kappa q}$ is the bare (unscreened) long-range Coulomb potential within each layer and $\kappa$ is the dielectric constant of the surrounding medium. The dynamical polarization bubble in the strained case is

\begin{equation}{\label{e2}}
\Pi_{\nu}(\textbf{q},i\omega) = -\frac{N}{16 v_{\nu x} v_{\nu y}}\frac{v_{\nu x}^2 q_x^2 + v_{\nu y}^2 q_y^2}{\sqrt{v_{\nu x}^2 q_x^2 + 
v_{\nu y}^2 q_y^2 + \omega^2}}
\end{equation}
with $N=4$ being the total number of fermion flavors, i.e. two spins and two valley degrees of freedom in graphene. On applying uniaxial strain within a given graphene sheet, assuming the limit of uniform bond deformations and neglecting all kinds of bond bending effects, the isotropic bandstructure of the Dirac fermions becomes anisotropic giving rise to two different velocities ($v_{\nu x}$ and $v_{\nu y}$) along the two spatial ($x-$ and $y-$) directions. With a fixed co-ordinate system and depending on the direction of the strain, as seen in Fig.~\ref{fig:2}, we define the anisotropy parameter proportional to the ratio of the two velocities such that the velocity along the direction of applied strain is always in the numerator. Thus if the strain is applied along the $x-$ ($y-$) direction of a given layer ($\nu$), then the anisotropy parameter is $v_{\nu \perp} = \frac{v_{\nu x}}{v_{\nu y}}$ ($v_{\nu \perp} = \frac{v_{\nu y}}{v_{\nu x}}$) where it is assumed $v_{\nu x} < v_{\nu y}$ ($v_{\nu y} < v_{\nu x}$) and that we are interested in the range $v_{\nu x}/v_{\nu y} \leq 1$ ($v_{\nu y}/v_{\nu x} \leq 1$). It is worth pointing out that in the numerical results to follow we'll approximate the larger velocity $v_{\nu y}$ ($v_{\nu x}$), for the applied strain along $x-$ ($y-$) direction respectively, equal to its isotropic limit $v$, i.e. the Fermi velocity.  This is due to the fact that in the tight-binding calculations of the uniaxially strained graphene, the larger velocity doesn't deviate much from its isotropic value\cite{pereira1}. For further details on the nature of anisotropic velocities and the bandstructure (Dirac cones) near a Dirac point due to the application of uniaxial strain in graphene, see for example Section II in Ref.~\onlinecite{sharma2}.\\
\indent The vdW force can be obtained from the interaction energy as

\begin{eqnarray}{\label{e3}}
\mathcal{F}(D) &=& -\frac{\textrm{d}\mathcal{E}(D)}{\textrm{d}D} \nonumber \\
&=& \frac{-2\hbar}{(2\pi)^{3}}\iint dq_{x} dq_{y} \int_{0}^{\infty} d\omega \hspace*{0.1cm} f(\textbf{q}, \omega ,D) \nonumber \\
\end{eqnarray}
where

\begin{equation}
f(\textbf{q},\omega ,D) = \frac{q e^{-2qD}\displaystyle  \prod_{\nu=1,2}V_{\nu}({\textbf{q}}) \Pi_{\nu}(\textbf{q},i\omega) 
}{\displaystyle\prod_{\nu=1,2}\epsilon_{\nu}(\textbf{q},i\omega) - e^{-2qD}\displaystyle\prod_{\nu=1,2}V_{\nu}({\textbf{q}}) \Pi_{\nu}(\textbf{q},i\omega)} \nonumber \\
\end{equation}
and it can be evaluated for each of the three different models considered in this work. A straightforward calculations gives

\begin{equation}{\label{e4}}
\mathcal{F}^{m}(D) = -\frac{\mathcal{C}^{m}(v_{\perp},\alpha)}{D^{4}} \\
\end{equation}
where $m$ = (a), (b) and (c) are the anisotropic models as shown in Fig.~\ref{fig:2} and the coefficient, $\mathcal{C}^{m}(v_{\perp},\alpha)$, is given by 

\begin{eqnarray}{\label{e5}}
\lefteqn{\mathcal{C}^{m}(v_{\perp},\alpha) =}  \nonumber \\
&& \frac{\hbar v}{8(2\pi)^3}\int_{0}^{2\pi} d\varphi \int_{0}^{\infty} d\tilde{\omega}  \int_{0}^{\infty} d\tilde{q} \left(\frac{\tilde{q}^3 e^{-\tilde{q}}}{g^{m}(\varphi, \tilde{\omega}, v_{\perp}, \alpha ) - e^{-\tilde{q}}}\right). \nonumber \\
\end{eqnarray}
We have used polar co-ordinates and considered dimensionless scaled variables for the frequency ($\tilde{\omega} = \frac{2D\omega}{v{\tilde{q}}}$) and the momentum (${\tilde{q}} = 2qD$) which sets the distance parameter (D) outside the integrand. The coefficient only depends on the strength of anisotropy ($v_{\perp}$) and the dimensionless intra-layer effective Coulomb coupling constant, $\alpha = \frac{e^2}{\kappa \hbar v}$. The function $g^{m}(\varphi, \tilde{\omega}, v_{\perp}, \alpha)$ depends explicitly on the way in which the uniaxial strain is applied.\\
\indent For the case when the anisotropy is only along one of the layers i.e., case (a) of Fig. 2, with $v_{1\perp} = 1$ and $v_{2\perp} = \frac{v_{2y}}{v} = v_{\perp}$, we have

\begin{eqnarray}{\label{e6}}
g^{(a)}(\varphi, \tilde{\omega}, v_{\perp}, \alpha) &&= \left( 1 + \frac{2\sqrt{1 + \tilde{\omega}^{2}}}{\pi \alpha} \right)\times \nonumber \\
&&\left( 1 + \frac{2v_{\perp} \sqrt{\cos^{2}{\varphi} + v_{\perp}^{2}\sin^{2}{\varphi} + \tilde{\omega}^{2}}}{\pi \alpha (\cos^{2}{\varphi} + v_{\perp}^{2}\sin^{2}{\varphi})} \right) \nonumber\\ 
\end{eqnarray} 

\indent On the other hand when equal amount of strain is applied on both the layers and along the same direction, i.e., case (b) with $v_{1\perp} = v_{2\perp} = v_{\perp}$, we have

\begin{equation}{\label{e7}}
g^{(b)}(\varphi, \tilde{\omega}, v_{\perp}, \alpha) = \left( 1 + \frac{2v_{\perp} \sqrt{\cos^{2}{\varphi} + v_{\perp}^{2}\sin^{2}{\varphi} + \tilde{\omega}^{2}}}{\pi \alpha (\cos^{2}{\varphi} + v_{\perp}^{2}\sin^{2}{\varphi})} \right)^2 
\end{equation}

And all the other limits, $v_{1\perp} \neq v_{2\perp}$, can also be studied but since these limits are intermediate between the models (a) and (b) therefore we don't consider them in this work.\\
\indent For the case (c), i.e., when both the layers are strained in different (perpendicular) directions we define $v_{1\perp} = \frac{v_{1x}}{v}$ and $v_{2\perp} = \frac{v_{2y}}{v}$. One can vary these two parameters independently but as before we consider $v_{1\perp} = v_{2\perp} = v_{\perp}$, which corresponds to maximum value of the force. In this case, the function $g^{m}(\varphi, \tilde{\omega}, v_{\perp}, \alpha)$ is given by

\begin{eqnarray}{\label{e8}}
g^{(c)}(\varphi, \tilde{\omega}, v_{\perp}, \alpha) &=& \left( 1 + \frac{2v_{\perp} \sqrt{v_{\perp}^{2}\cos^{2}{\varphi} + \sin^{2}{\varphi} + \tilde{\omega}^{2}}}{\pi \alpha (v_{\perp}^{2}\cos^{2}{\varphi} + \sin^{2}{\varphi})} \right) \nonumber \\
&&\!\!\!\!\!\times\left( 1 + \frac{2v_{\perp} \sqrt{\cos^{2}{\varphi} + v_{\perp}^{2}\sin^{2}{\varphi} + \tilde{\omega}^{2}}}{\pi \alpha (\cos^{2}{\varphi} + v_{\perp}^{2}\sin^{2}{\varphi})} \right) \nonumber \\
\end{eqnarray} 
\indent Before we discuss the numerical results of the vdW force coefficient for three anisotropic models, we confirm the isotropic limit. With $v_{1\perp} = v_{2\perp} = v_{\perp} = 1$, $\hbar = 6.58 \times 10^{-16} $ eV s, $v = 10^{16}$ {\AA}  $s^{-1}$, we get $\mathcal{C}(1,\alpha) =$ 0.40 eV {\AA} for the freely suspended ($\kappa = 1$ and $\alpha = 2.2$) graphene sheets. This is the well-established value for the force in isotropic graphene\cite{gomezsantos,klimchitskaya}.\\

\section{\label{sec:numres}Evolution of the van der Waals force with strain and electron-electron interaction strength}

\indent In this section we present the results for the vdW force coefficient for varying strength of anisotropy and effective coupling constant. We numerically evaluate the coefficient for three anisotropic models, shown in Fig.~\ref{fig:2}, by substituting Eq.~(\ref{e6}),~(\ref{e7}) and ~(\ref{e8}) in Eq.~(\ref{e5}) for models (a), (b) and (c) respectively. First we shall carry out the analysis
for fixed (momentum independent) values of the couplings (Section~\ref{sec:woeei}). Then in Section~\ref{sec:weei} we take into account the logarithmic  coupling renormalization due to intra-layer electron-electron interactions. \\

\subsection{\label{sec:woeei} Results for the van der Waals force} 

\begin{figure}
\centering
\includegraphics[height=7.0cm,width=9.0cm]{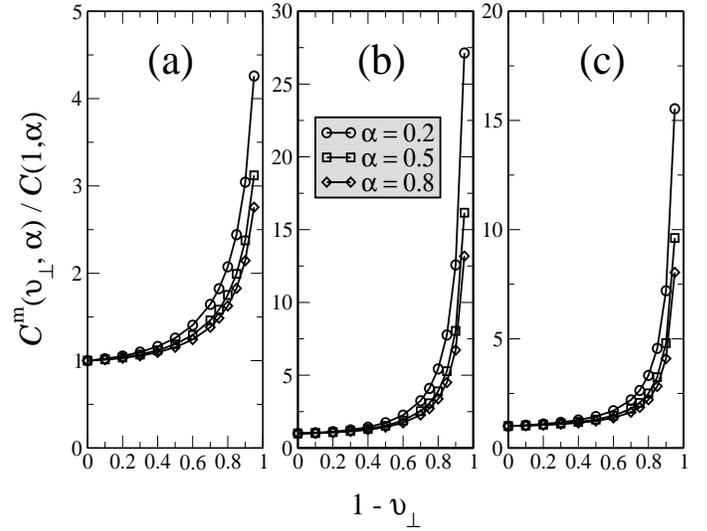}
\caption{\label{fig:3} (Color online) The ratio of the anisotropic, $\mathcal{C}^{m}(v_{\perp},\alpha)$, to the isotropic, $\mathcal{C}($1$,\alpha)$, coefficient of the vdW force as a function of variation in the anisotropy (1 - $v_{\perp}$) and strength of coupling constant ($\alpha$) for three different models.}
\end{figure}

\indent Let us begin by calculating  the vdW force for fixed (i.e. momentum-independent, unrenormalized) values of the model parameters, $v_{\perp}$ and $\alpha$. In Fig.~\ref{fig:3} we plot the ratio of the anisotropic, $\mathcal{C}^{m}(v_{\perp},\alpha)$ , to the isotropic, $\mathcal{C}($1$,\alpha)$ , coefficient as a function of change in anisotropy parameter (1 - $v_{\perp}$) and varying strength of effective coupling constant ($\alpha$) for different models as shown in three panels in the figure. For our calculations, we consider three different values of the effective interaction, $\alpha$ = 0.2, 0.5 and 0.8 shown in circle, square and diamond respectively in the figure and for these values we get the isotropic coefficient as $\mathcal{C}($1$,\alpha)$ = 0.014, 0.059 and 0.115 eV {\AA} respectively.\\
\indent As seen in Fig.~\ref{fig:3}, for all the anisotropic models we find that for fixed coupling constant the coefficient increases with increasing anisotropy which can be understood as an increase in the intra-layer charge susceptibility.  From the three anisotropic models considered in this work, the effect of strain and interaction is the most prominent for model (b) which is the case when both the layers are strained in the same direction and where we notice an order of magnitude increase in the vdW coefficient at the maximum value of applied strain. We also observe that at any given finite value of strain, the factor of increase in the strength of the force of attraction gets reduced with increasing coupling. \\
\indent We restrict our calculations to the maximum anisotropy of $v_{\perp}$ = 0.05. In the limit of large applied uniaxial strain ($v_{\perp} \rightarrow 0$) in any of the anisotropic models, it is straightforward to see that the vdW force coefficient diverges due to the integration over the frequency. This suggests that  the retardation-induced upper cutoff\cite{gomezsantos} for the scaled dimensionless frequency should be kept, i.e the integration should be  performed as $\int_{0}^{c/v} d \tilde{\omega}$. This accounts for the retardation effect  where $c$ is the speed of light in vacuum and $c/v=300$;  the non-retarded results are obtained in the limit $c \rightarrow \infty$. We have performed all of our calculations with the cutoff as described above. It is known that in isotropic graphene the introduction of this cutoff has practically no effect on the results \cite{gomezsantos}. However we find that  in the presence of applied strain when the polarization and thus the force increase substantially, retardation effects gain importance in the limit of large separation between the two graphene layers.\\

\subsection{\label{sec:weei} Influence of many-body logarithmic renormalization} 

\indent The intriguing many-body physics due to the electron-electron interaction is known to renormalize the Fermi velocity ($v$) and the effective coupling constant ($\alpha$) in graphene\cite{kotov}. In the strained case the anisotropic parameter ($v_{\perp}$), which depends on the Fermi velocity, also gets altered and acquires logarithmic corrections due to the many-body interactions\cite{sharma1}. \\
\indent Recently\cite{dobson1}, the effect of such renormalization was studied in the context of vdW force between two undoped and unstrained graphene layers where the authors concluded that the asymptotic behavior of the vdW interaction at large separation gets modified due to change in the low-energy (long wavelength) behavior of the Dirac fermions in the presence of electron-electron interactions. They also showed that not only the vdW force coefficient but also the power law dependence on the distance between the two graphene layers could drastically change.\\
\indent We have included the many-body interactions and re-calculated the vdW force coefficient which gets modified due to the renormalization of the physical quantities ($v(l), \alpha(l), v_{\perp}(l)$), where $l$ is the logarithmic scale in the renormalization group treatment (see below). Their behavior is governed, to lowest order in the interaction (i.e. under the condition $\alpha \ll 1$), by the following coupled nonlinear renormalization group (RG) equations\cite{sharma1}:

\begin{equation}{\label{e9}}
\frac{\textrm{d}v}{\textrm{d}l} = \frac{v\alpha}{4\pi} \int_{0}^{2\pi} d\varphi\frac{\cos^2{\varphi}}{\sqrt{\cos^2{\varphi} + v^2_{\perp} \sin^2{\varphi}}}
\end{equation}
\begin{equation}{\label{e10}}
\frac{\textrm{d}\alpha}{\textrm{d}l} = -\frac{\alpha^2}{4\pi} \int_{0}^{2\pi} d\varphi\frac{\cos^2{\varphi}}{\sqrt{\cos^2{\varphi} + v^2_{\perp} \sin^2{\varphi}}}
\end{equation}
\begin{equation}{\label{e11}}
\frac{\textrm{d}v_{\perp}}{\textrm{d}l} = \frac{v_{\perp}\alpha}{4\pi} \int_{0}^{2\pi} d\varphi\frac{\sin^2{\varphi}-\cos^2{\varphi}}{\sqrt{\cos^2{\varphi} + v^2_{\perp} \sin^2{\varphi}}}
\end{equation}
where $l = \ln(\Lambda/q) = \ln{(2 D \Lambda/\tilde{q})}$ is the RG parameter with $\Lambda$ ($\sim 1/a$) being the momentum cutoff and $a = 1.42${\AA} is the lattice constant of graphene. Our results are valid 
in the weak coupling regime, which is a good approximation since $\alpha(l)$ decreases under renormalization.\\

\begin{figure}
\centering
\includegraphics[height=7.0cm,width=9.0cm]{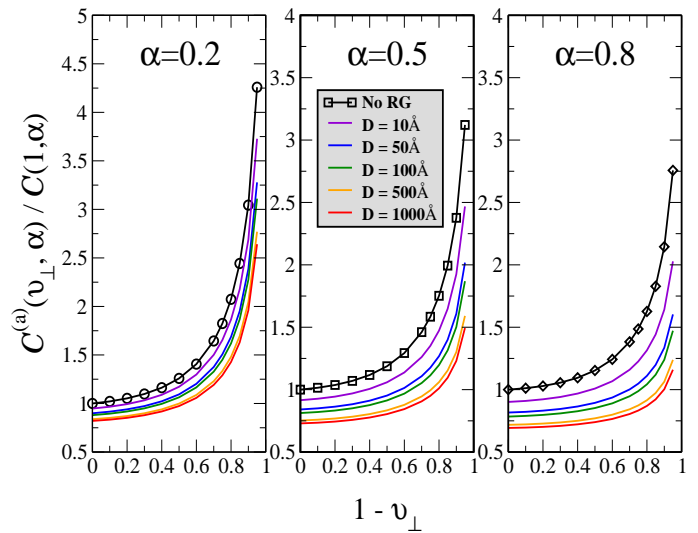}
\caption{\label{fig:4} (Color online) The ratio of anisotropic, $\mathcal{C}^{(a)}(v_{\perp},\alpha)$, to the isotropic, $\mathcal{C}($1$,\alpha)$, coefficient of the vdW force between two graphene layers for the anisotropic model (a) and with many-body effects. The renormalization of the Fermi velocity is considered for different values of initial anisotropy and coupling constant which depend on the separation between the two graphene layers.}
\end{figure}

\indent One of the consequences of RG is that in Eq.~(\ref{e5}), the function $g^{m}(\varphi\tilde{\omega}, v_{\perp}, \alpha)$ becomes distance and momentum dependent, i.e. $g^{m}(\varphi, \tilde{\omega}, v_{\perp}(l), \alpha(l))$. Since this dependence makes it difficult to carry out the momentum integration in Eq.~(\ref{e5}), we can  perform it approximately by setting $\tilde{q} = 3$ under the logs. This is justified using the fact that the numerator in  Eq.~(\ref{e5}) is a rapidly decreasing function of $\tilde{q}$ and has its extremum at $\tilde{q} \sim 3$ while the denominator contains $v_{\perp}(l)$ and $\alpha(l)$ which are logarithmically slow varying functions of momentum since they depend on the RG scale $l$. In order words in Eq.~(\ref{e5}) we can approximate:
\begin{equation*}
\int_{0}^{\infty} d\tilde{q}
 \frac{\tilde{q}^3e^{-\tilde{q}}}{g^{m}(\varphi,\tilde{\omega},v_{\perp}(\ln{(2\Lambda D/\tilde{q})}),
\alpha(\ln{(2\Lambda D/\tilde{q})}))  -e^{-\tilde{q}}} \approx
\end{equation*}
\begin{equation}
\int_{0}^{\infty} d\tilde{q}
 \frac{\tilde{q}^3e^{-\tilde{q}}}{g^{m}(\varphi,\tilde{\omega},v_{\perp}(\ln{(2\Lambda D/3}),
\alpha(\ln{(2\Lambda D/3)})) -e^{-\tilde{q}}} 
\end{equation}
Moreover, we have numerically checked the validity of our approximation for the isotropic case where the RG equations can be analytically solved thereby providing a possibility to make a comparison between the exact and the approximate calculation.\\

\begin{figure}
\centering
\includegraphics[height=7.0cm,width=9.0cm]{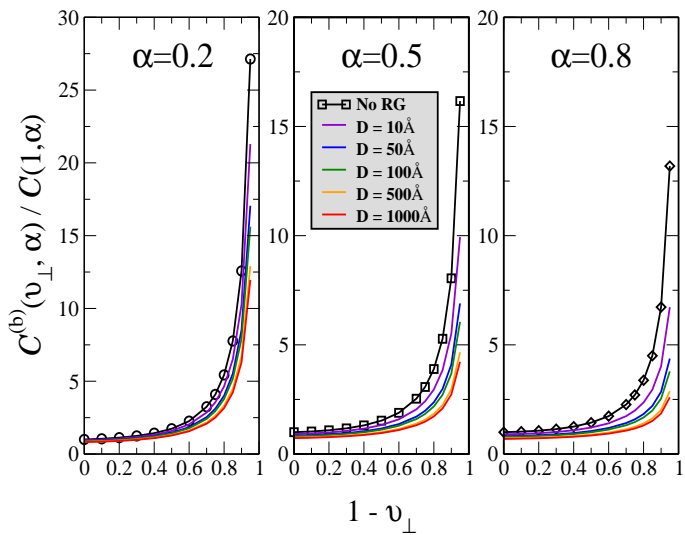}
\caption{\label{fig:5} (Color online) The same as in Fig.~\ref{fig:4} but for model (b).}
\end{figure}

\begin{figure}
\centering
\includegraphics[height=7.0cm,width=9.0cm]{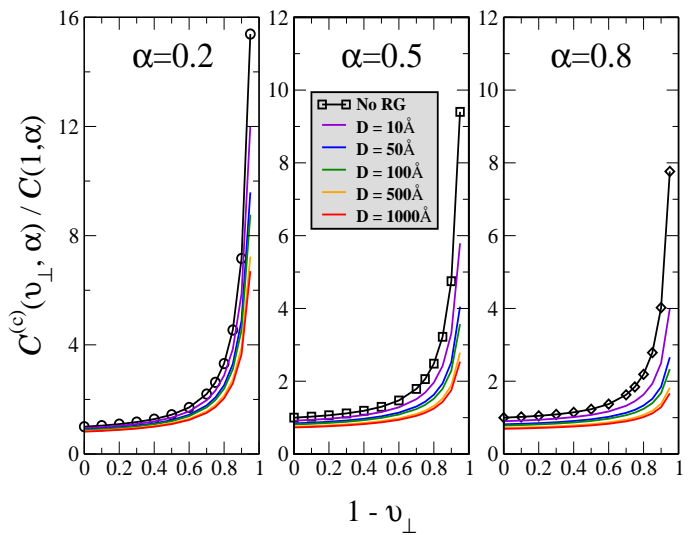}
\caption{\label{fig:6} (Color online) The same as in Fig.~\ref{fig:4} but for model (c).}
\end{figure}
\indent With this assumption, we evaluate the vdW force coefficient as a function of model parameters but with the additional dependence  on distance (D) of  the anisotropy $v_{\perp}(\ln{(2 D \Lambda/3)})$ and the coupling constant  $\alpha(\ln{(2 D \Lambda/3)})$. The results are plotted in Figures \ref{fig:4}, \ref{fig:5} and \ref{fig:6} for model (a), (b) and (c) respectively, for different values of the separation between the graphene layers.\\
\indent We first evaluate Eq.~(\ref{e5}) for the case described by Eq.~(\ref{e6}),  i.e. model (a), along with Eqs.~(\ref{e9}--\ref{e11}), for three different values of coupling constants and for various distances between the graphene layers. We plot the ratio of the renormalized anisotropic to the unrenormalized isotropic coefficient as shown in Fig.~\ref{fig:4}. For comparison purpose, we also include the results without many-body interactions (no RG) as taken from Fig.~\ref{fig:3}. As the distance between the two layers increases there is an overall decrease in the vdW force coefficient which is seen even for the isotropic case. This happens because in the limit of large separation the low energy physics becomes important and the renormalization effect is the strongest at that energy scale. It is known \cite{sharma1} that the interplay between the weak coupling behavior and anisotropy in the presence of electron-electron interactions is such that the RG flow is towards the non-interacting and isotropic limit. Therefore as the distance between the graphene layers is increased (and thus the RG scale increases) the many-body renormalization effects become more pronounced  resulting in the decrease of the effective interaction $\alpha(l)$ which in turn leads to the decrease of the vdW force. \\

\begin{figure}
\centering
\includegraphics[height=7.0cm,width=9.0cm]{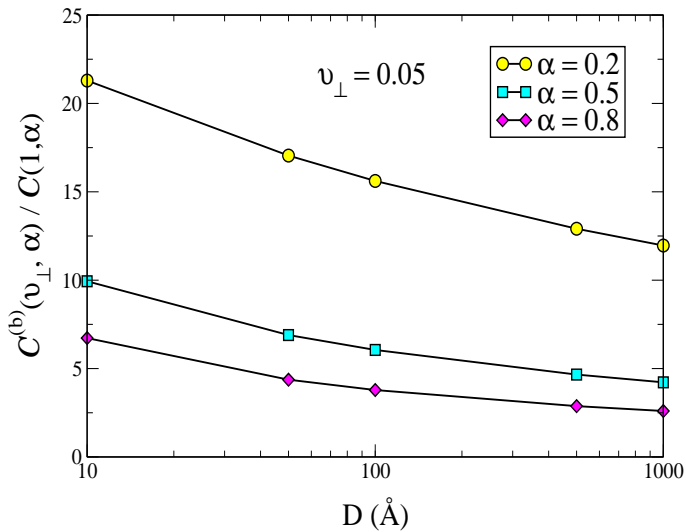}
\caption{\label{fig:7} (Color online) The ratio of anisotropic, $\mathcal{C}^{(b)}(v_{\perp},\alpha)$, to the isotropic, $\mathcal{C}($1$,\alpha)$, coefficient of the vdW force between two graphene layers, for model (b), as a function of variation in the strength of coupling constant,$\alpha$, and the distance between the layers for fixed anisotropy, $v_{\perp}$=0.05.}
\end{figure}

\indent Such feature is also found in other anisotropic models but the dominant effect is seen for model (b) which is the case where the strain is applied on both the layers in the same direction. In Fig.~\ref{fig:7} we plot the ratio of anisotropic to the isotropic coefficient for model (b) as a function of variation in the strength of coupling constant, $\alpha$, and the distance between the layers for fixed anisotropy, $v_{\perp}$=0.05, which is the maximum value of strain considered in this work. We see that the decrease in the coefficient is rapid for the weakly interacting case ($\alpha = 0.2$) which validates our consideration of  weak electron-electron interactions.\\

\section{\label{sec:sumcon} Summary and Conclusions}

\indent In this work  we studied the vdW force acting between two uniaxially strained undoped graphene layers at zero temperature and separated by a finite distance. We evaluated the vdW force coefficient as a function of anisotropy due to strain and for different values of the Coulomb coupling constant. In order to study the anisotropy we considered three different models: (a) where one of the two layers is uniaxially strained, (b) the two layers are strained in the same direction, and (c) one of the layers is strained in the perpendicular direction with respect to the other. We calculated the vdW interaction energy and the vdW force per unit area in each of these cases. Our analysis was based on the effective long-range Coulomb interaction between the graphene sheets derived within the Random Phase Approximation (RPA) since it is known that this approach is equivalent to the Lifshitz approach with weak retardation. We also included the dominant retardation effect which manifests itself as an effective frequency cutoff. In fact we found that for large strain, where the force grows dramatically due to the increased polarization, retardation provides an effective suppression of this growth and its inclusion is essential. \\
\indent We performed our calculations by first neglecting effects arising from renormalization of electron-electron interactions, i.e. running of the couplings. We found that for all three anisotropic models and for any given strength of the coupling, the vdW force increases with increasing anisotropy.  The effect is most prominent for the case when both  layers were strained in the same direction. In this case we observed up to an order of magnitude increase in the strained relative to the unstrained case. We also found that at any given finite value of strain, the increase of the vdW force is suppressed as the coupling increases. \\
\indent These calculations were followed by an investigation which included the effect of electron electron interaction renormalization in the region of large separation between the strained graphene layers. We found that the many-body renormalization contributions to the correlation energy were non negligible and the vdW interaction energy decreased as a function of increasing distance between the layers due to renormalization of the Fermi velocity, the effective interaction, and the anisotropy. Our study can be helpful in designing modern graphene-based vdW heterostructures which have attracted a lot of attention in recent research activities. Strong anisotropies leading to enhanced vdW attraction could be potentially achieved in artificial graphene-based lattices. Moreover, interactions between multilayer structures containing more than two sheets, non-zero doping and finite temperature could be studied in the future work. \\ 

\section{\label{sec:ack} Acknowledgments} 

We are grateful to D. Clougherty and A. Del Maestro for numerous stimulating discussions on the subject of dispersion forces. The research of A. Sharma, V. N. Kotov and A. H. Castro Neto was supported by the U.S. Department of Energy (DOE) grant DE-FG02-08ER46512. AHCN acknowledges the NRF-CRP award "Novel 2D materials with tailored properties: beyond graphene" (R-144-000-295-281). A. Sylvester acknowledges financial assistance from the Research Experiences for Undergraduates (REU) Program of the National Science Foundation (No. DMR-1062966).

\end{document}